\documentclass[twocolumn,conference]{IEEEtran}
\usepackage[left=1in, right=1in, bottom=1in, top=1in]{geometry}

\usepackage[utf8]{inputenc}
\usepackage[T1]{fontenc}
\usepackage[scaled=0.92]{helvet}  
\usepackage[noadjust]{cite}

\usepackage{graphicx}
\usepackage{array}
\newcolumntype{M}[1]{>{\centering\arraybackslash}m{#1}}
\usepackage{tabularx}
\usepackage{dblfloatfix}
\usepackage{amsmath}
\usepackage{newtxtext, newtxmath}
\usepackage{dsfont}
\usepackage{hhline,booktabs}
\usepackage{tablefootnote}
\usepackage{enumitem}
\usepackage[T1]{fontenc}
\usepackage{adjustbox}
\usepackage{booktabs, multirow}
\usepackage{amsmath}

\usepackage[scaled=0.9]{DejaVuSansMono}
\usepackage{listings}
\usepackage{amsfonts}

\usepackage{amsthm,thmtools,xcolor}
\declaretheoremstyle[
  headfont=\color{red}\normalfont\bfseries,
  bodyfont=\color{red}\normalfont\itshape,
]{colored}

\usepackage{bm}
\usepackage{url}
\usepackage{mathtools}
\usepackage{balance}
\usepackage{subcaption}
\allowdisplaybreaks

\usepackage{algorithm}
\usepackage[noend]{algpseudocode}
\usepackage{etoolbox}
\usepackage[mathscr]{euscript}
\usepackage{calc}
\usepackage{pgfplots}
\usepackage{tikz}
\usepackage{setspace}
\usepackage{pgfplotstable}

\usepackage[font=scriptsize]{caption}
\DeclareMathAlphabet{\pazocal}{OMS}{zplm}{m}{n}
\DeclareSymbolFont{missing}{OML}{cmr}{m}{n}
\DeclareMathSymbol{\ell}{\mathord}{missing}{'140}

\usetikzlibrary{calc}
\usetikzlibrary{patterns}
\usetikzlibrary{spy,backgrounds}
\pgfplotsset{grid style={dotted,gray}}

\newcommand{\maximize}{%
  \mathopen{}\operatorname*{maximize}%
}

\newcommand{\subjto}{\textup{subject to}}

\newcounter{problem}
\newcounter{save@equation}
\newcounter{save@problem}

\newlength{\depthofsumsign}
\newlength{\totalheightofsumsign}
\newlength{\heightanddepthofargument}

\makeatletter

\tikzset{reset label anchor/.code={%
    \let\tikz@auto@anchor=\pgfutil@empty
    \def\tikz@anchor{#1}
  },
  reset label anchor/.default=center
}

\let\save@mathaccent\mathaccent
\newcommand*\if@single[3]{%
  \setbox0\hbox{${\mathaccent"0362{#1}}^H$}%
  \setbox2\hbox{${\mathaccent"0362{\kern0pt#1}}^H$}%
  \ifdim\ht0=\ht2 #3\else #2\fi
}
\newcommand*\rel@kern[1]{\kern#1\dimexpr\macc@kerna}
\newcommand*\widebar[1]{\@ifnextchar^{{\wide@bar{#1}{0}}}{\wide@bar{#1}{1}}}
\newcommand*\wide@bar[2]{\if@single{#1}{\wide@bar@{#1}{#2}{1}}{\wide@bar@{#1}{#2}{2}}}
\newcommand*\wide@bar@[3]{%
  \begingroup
  \def\mathaccent##1##2{%
    \let\mathaccent\save@mathaccent
    \if#32 \let\macc@nucleus\first@char \fi
    \setbox\z@\hbox{$\macc@style{\macc@nucleus}_{}$}%
    \setbox\tw@\hbox{$\macc@style{\macc@nucleus}{}_{}$}%
    \dimen@\wd\tw@
    \advance\dimen@-\wd\z@
    \divide\dimen@ 3
    \@tempdima\wd\tw@
    \advance\@tempdima-\scriptspace
    \divide\@tempdima 10
    \advance\dimen@-\@tempdima
    \ifdim\dimen@>\z@ \dimen@0pt\fi
    \rel@kern{0.6}\kern-\dimen@
    \if#31
    \overline{\rel@kern{-0.6}\kern\dimen@\macc@nucleus\rel@kern{0.4}\kern\dimen@}%
    \advance\dimen@0.4\dimexpr\macc@kerna
    \let\final@kern#2%
    \ifdim\dimen@<\z@ \let\final@kern1\fi
    \if\final@kern1 \kern-\dimen@\fi
    \else
    \overline{\rel@kern{-0.6}\kern\dimen@#1}%
    \fi
  }%
  \macc@depth\@ne
  \let\math@bgroup\@empty \let\math@egroup\macc@set@skewchar
  \mathsurround\z@ \frozen@everymath{\mathgroup\macc@group\relax}%
  \macc@set@skewchar\relax
  \let\mathaccentV\macc@nested@a
  \if#31
  \macc@nested@a\relax111{#1}%
  \else
  \def\gobble@till@marker##1\endmarker{}%
  \futurelet\first@char\gobble@till@marker#1\endmarker
  \ifcat\noexpand\first@char A\else
  \def\first@char{}%
  \fi
  \macc@nested@a\relax111{\first@char}%
  \fi
  \endgroup
}

\newenvironment{problem}
{\setcounter{problem}{\value{save@problem}}%
  \setcounter{save@equation}{\value{equation}}%
  \let\c@equation\c@problem
  \subequations
}
{\endsubequations
  \setcounter{save@problem}{\value{equation}}%
  \setcounter{equation}{\value{save@equation}}%
}

\algnewcommand{\LineComment}[1]{\Statex \hskip\ALG@thistlm
  \(\triangleright\) #1}
\def\BState{\State\hskip-\ALG@thistlm}

\algdef{SE}[DOWHILE]{Do}{EndDoWhile}{\algorithmicdo}[1]{\algorithmicwhile\ #1}%

   \tikzset{nomorepostaction/.code=\let\tikz@postactions\pgfutil@empty}

   \long\def\ifnodedefined#1#2#3{%
   \@ifundefined{pgf@sh@ns@#1}{#3}{#2}%
 }
\usetikzlibrary{decorations.pathmorphing,arrows.meta,positioning}
\tikzstyle{printersafe}=[decoration={amplitude=0pt}]

\makeatother

\usetikzlibrary{automata}
\usetikzlibrary{calc,fit,shapes.geometric}
\pgfdeclarelayer{signal}
\pgfsetlayers{signal,main}
\tikzstyle{printersafe}=[segment amplitude=0 pt]
\usetikzlibrary{arrows}
\newcounter{cntr}
\usetikzlibrary{calc}
\usetikzlibrary{decorations.pathreplacing,decorations.markings,shapes.geometric}
\tikzset{naming/.style={align=center}}
\tikzset{antenna/.style={insert path={-- coordinate (ant#1) ++(0,0.5) -- +(135:0.5) + (0,0) -- +(45:0.5)}}}
\tikzset{station/.style={naming,draw,shape=dart,shape border rotate=90, minimum width=20mm, minimum height=20mm,outer sep=0pt,inner
    sep=3pt}}
\tikzset{mobile/.style={naming,draw,shape=rectangle,minimum width=12mm,minimum height=6mm, outer sep=0pt,inner sep=3pt}}
\tikzset{radiation/.style={{decorate,decoration={expanding waves,angle=90,segment length=6pt}}}}

\tikzset{
  every pin/.style={rectangle,rounded corners=3pt,font=\footnotesize},
  small dot/.style={fill=black,circle,scale=0.5}
}

\tikzset{
  invisible/.style={opacity=0},
  visible on/.style={alt={#1{}{invisible}}},
  alt/.code args={<#1>#2#3}{%
    \alt<#1>{\pgfkeysalso{#2}}{\pgfkeysalso{#3}} 
  },
}

\tikzset{pics/.cd,
  SBS/.style={code={
      \begin{scope}[local bounding box=#1]
        \fill [pic actions/.try] (-1,0) -- (-1/2,3) -- (1/2, 3) -- (1,0) -- cycle;
        \fill [pic actions/.try] (-1/16,2) rectangle (1/16,4);
        \fill [pic actions/.try] (0,4) circle [radius=1/4];
        \foreach \i in {-1,1}
        \fill [shift=(90:4), xscale=\i]
        \foreach \r in {1,3/2,2}{
          (-45:\r) arc (-45:45:\r) -- (45:\r-1/10)
          arc(45:-45:\r-1/10) -- cycle
        };
      \end{scope}
    }},
  MBS/.style={code={
      \begin{scope}[local bounding box=#1]
        \fill [pic actions/.try] (-1,0) -- (-1/2,3) -- (1/2, 3) -- (1,0) -- cycle;
        \fill [pic actions/.try] (-1/16,2) rectangle (1/16,4);
        \fill [pic actions/.try] (0,4) circle [radius=1/4];
        \foreach \i in {-1,1}
        \fill [shift=(90:4), xscale=\i]
        \foreach \r in {1,3/2,2}{
          (-45:\r) arc (-45:45:\r) -- (45:\r-1/10)
          arc(45:-45:\r-1/10) -- cycle
        };
      \end{scope}
    }},
  SU/.style={code={
      \begin{scope}[local bounding box=#1]
        \fill [even odd rule, pic actions/.try]
        (-1,-5/2) -- (-1,-1/8) -- (1,-1/8) -- (1,-5/2)
        arc (360:180:1 and 1/4) -- cycle (-1,5/2) -- (-1,1/8) -- (1,1/8) -- (1,5/2)
        arc (0:180:1 and 1/4) -- cycle (-3/4, 9/4) -- (-3/4, 3/8) -- (3/4, 3/8) -- (3/4, 9/4)
        arc (0:180:3/4 and 1/8)-- cycle
        \foreach \i in {-1,0,1}{\foreach \j in {1,2,3}{
            (-\i*1/2-3/16,-\j/2-3/4) rectangle ++(3/8, 3/8)
          }
        }
        (-1/2,-3/4) rectangle (1/2, -1/4);
      \end{scope}
    }},
  MU/.style={code={
      \begin{scope}[local bounding box=#1]
        \fill [even odd rule, pic actions/.try]
        (-1,-5/2) -- (-1,-1/8) -- (1,-1/8) -- (1,-5/2)
        arc (360:180:1 and 1/4) -- cycle (-1,5/2) -- (-1,1/8) -- (1,1/8) -- (1,5/2)
        arc (0:180:1 and 1/4) -- cycle (-3/4, 9/4) -- (-3/4, 3/8) -- (3/4, 3/8) -- (3/4, 9/4) arc (0:180:3/4 and 1/8)-- cycle
        \foreach \i in {-1,0,1}{
          \foreach \j in {1,2,3}{
            (-\i*1/2-3/16,-\j/2-3/4) rectangle ++(3/8, 3/8)
          }
        }
        (-1/2,-3/4) rectangle (1/2, -1/4);
      \end{scope}
    }},
  SIGNAL/.style={code={
      \begin{scope}[local bounding box=#1]
        \fill [pic actions/.try]
        (0,-3) -- (-1,1/2) -- (1/8,1/4) -- (0,3) -- (1,-1/2) -- (-1/8,-1/4) -- cycle;
      \end{scope}
    }},
  queuei/.style={code={
      \begin{scope}
        \stepcounter{cntr}
        \node[inner sep=0pt, outer sep=0pt,draw,rectangle split,rectangle split horizontal,minimum height=0.5cm,rectangle split parts=3]
        (queue-\thecntr) [pic actions] {};
        \draw
        (queue-\thecntr.north west) -- ++(-0.2cm,0)
        (queue-\thecntr.south west) -- ++(-0.2cm,0);
        \node[above] at ([xshift=-0.5cm]queue-\thecntr.north)
        {$Q_#1$};
      \end{scope}
    }}
}
\colorlet{sky blue}{blue!60!cyan!75!black}
\colorlet{dark blue}{blue!50!cyan}
\colorlet{chameleon}{olive!75!green}
\tikzset{signal/.style={->,draw=black, line width=0.05em, dashed,printersafe}}

\newsavebox{\mybox}
\pgfplotsset{compat=1.16}
\begin{document}

\title{Competitive Algorithms and Reinforcement Learning for NOMA in IoT Networks}

\author{\IEEEauthorblockN{Zoubeir~Mlika,~and Soumaya~Cherkaoui}
\IEEEauthorblockA{Department of Electrical and Computer Engineering, University of Sherbrooke\\ 
zoubeir.mlika@usherbrooke.ca, soumaya.cherkaoui@usherbrooke.ca}}%

\maketitle

\begin{abstract}
  This paper studies the problem of massive Internet of things (IoT) access in beyond fifth generation (B5G) networks using non-orthogonal multiple access (NOMA) technique. The problem involves massive IoT devices grouping and power allocation in order to respect the low latency as well as the limited operating energy of the IoT devices. The considered objective function, maximizing the number of successfully received IoT packets, is different from the classical sum-rate-related objective functions. The problem is first divided into multiple NOMA grouping subproblems. Then, using competitive analysis, an efficient online competitive algorithm (CA) is proposed to solve each subproblem. Next, to solve the power allocation problem, we propose a new reinforcement learning (RL) framework in which a RL agent learns to use the CA as a black box and combines the obtained solutions to each subproblem to determine the power allocation for each NOMA group. Our simulations results reveal that the proposed innovative RL framework outperforms deep-Q-learning methods and is close-to-optimal.
\end{abstract}

\begin{IEEEkeywords}
  Internet of things, non-orthogonal multiple access, online grouping, online power allocation, online competitive algorithms, reinforcement learning.
\end{IEEEkeywords}

\newcommand{\describeContent}[1]{%
\begingroup%
\let\thefootnote\relax%
\footnotetext{#1}%
\endgroup%
}

\IEEEpeerreviewmaketitle


\section{Introduction}\label{section:introduction}
Internet of things (IoT) will soon be composed of tens of billions of objects that are connected to the Internet and that communicate with each other without (or with little) human interactions~\cite{8861118,ndih2016enhancing,7835337}. In a cellular-based IoT network, a massive number of objects (or, interchangeably, devices) can communicate with each other through the cellular network infrastructure~\cite{7736615}, e.g., a next generation nodeB (gNB). The massive access problem in IoT networks is thus expected to become a challenging problem for beyond fifth generation (B5G) networks. The problem is even more challenging when IoT devices~\cite{7132717} operate with limited energy and require communications with very low latency.

In~\cite{8663999}, the authors consider the problem of uplink grant in machine-to-machine networks. The problem is transformed into predicting which IoT device has packets to send and a two-stage machine learning solution is developed. In~\cite{8644350}, the problem of fast uplink grant access is solved based on a multi-armed bandit approach. The objective is to maximize a utility function that is a combination of data rate, access delay, and value of data packets. A sleeping multi-armed bandit technique is used to model the situation where the set of possible actions is not known in advance. In~\cite{8365765}, the authors study the problem of maximizing the number of served IoT devices using the non-orthogonal multiple access (NOMA). The power allocation is first obtained by solving the feasibility problem of minimum rate requirements. Then, the NOMA channel assignment  is solved by reducing it to a maximum independent set problem. Other related works solve the problem using off-the-shelf deep reinforcement learning (DRL) approaches~\cite{9145187,8790780,9174918,abouaomar2021deep}. In~\cite{9145187}, the authors solve the uplink NOMA user clustering problem in IoT networks. The proposed algorithm performs user clustering based on network traffic and it uses SARSA-based deep-Q-learning (DQL) method in the case of light network traffic and DRL in case of heavy traffic. In~\cite{8790780}, the authors solve the problem of power allocation and channel assignment in wireless networks using attention-based neural network. An optimization framework is first proposed to obtain the optimal power allocation. Then, a DRL framework is proposed to learn the channel assignment. In~\cite{9174918}, the authors propose to improve the random access channel procedure in real-time using DRL methods. Then, a decoupled learning algorithm is proposed based on recurrent neural network to train DRL agents. The majority of the works optimize sum-rate-related objectives and use DQL methods to solve the problem. To the best of our knowledge, this work is the first to prove that online competitive algorithms (CAs) can meet classical RL methods to outperform DQL without requiring huge training-intensive tasks. In~\cite{abouaomar2021deep}, the authors studied the service migration in network function virtualization of multi-access edge computing-based vehicular networks and solved the problem using deep Q learning.

To address the problem of network access by an increasingly big number of IoT devices, the NOMA technique was proposed for B5G networks~\cite{9205230}. In this paper, we study the \textbf{online}\footnote{An online problem is when its input is not available from the start but it is revealed one by one without knowing its future values.} grouping and power allocation problem in an IoT B5G network using NOMA. It is shown in~\cite{9383093} that maximizing the sum-rate can be achieved with few served devices. Thus, contrary to most previous works that maximize sum-rate-related objective functions, we focus on maximizing the number of successfully received packets. To fill this research gap, we propose a new RL framework called competitive reinforcement learning (CRL). The proposed framework is inspired by the divide-and-conquer technique and works mainly as follows. First, CRL starts by solving a special case of the considered problem, the online NOMA grouping problem, using online CAs. Next, a RL agent uses this CA as a black box to learn its optimal policy regarding to the power allocation of each NOMA group. The simulation results reveal that the proposed innovative RL framework outperforms DQL methods and is close-to-optimal.

Developing online algorithms for this problem is a challenging task. Our contributions are summarized as follows.
  We first give a mathematical programming model to solve the problem in an \textbf{offline} manner using off-the-shelf solvers. This part is important as it provides upper bounds for our \textbf{online} solutions or for future research improvement solutions.
  Then, we propose an online algorithm based on competitive algorithms as well as on reinforcement learning methods~\cite{9383093,abouaomar2021service} and we show that the proposed method is very simple as it does not require training-intensive tasks and it outperforms classical DQL methods~\cite{mlika2021network}.

The paper is organized as follows. Section~\ref{sec:model} presents the system model and formulates the problem. Section~\ref{sec:learn} presents the proposed CRL framework. Section~\ref{sec:simulations} illustrate some results, and finally, section~\ref{sec:conclusions} provides some insights and conclusions.

\section{System Model}\label{sec:model}
A cellular-based IoT network is considered in which there are one gNB and $M$ devices. Time is discrete and divided into $T$ frames where each frame is composed of $N$ slots. In each frame $t$, device $i$ has a packet of length (in bits) $L_{i}(t)\geqslant0$ to send. (In general, $L_{i}(t)$ may be zero for some $t$ in which case $i$ has no packet to send in frame $t$.) Device $i$'s packet in frame $t$ has a time of arrival and a deadline which are denoted by $a_{i}(t)$ and $d_{i}(t)$, respectively. The considered frame structure and the traffic pattern are similar to the frame-synchronized traffic pattern~\cite{8482322}. During the whole time horizon of $T$ frames, device $i$ has $p_{i}^{\textsf{max}}$ units of energy (or, without loss of generality, power) stored in its battery. A resource block (RB) is denoted by the pair $(j, t)$ for slot $j$ of frame $t$ and has a bandwidth of $W$ Hz. 

The channel gain between device $i$ and the gNB over RB $(j,t)$ is given by $h_{ij}(t)$. Let $x_{ij}(t)=1$ if and only if device $i$ transmits its packet using RB $(j,t)$ and let $p_{ij}(t)$ denote the transmission power of device $i$ using RB $(j,t)$. The signal to interference-plus-noise ratio (SINR) of device $i$ at the gNB using RB $(j,t)$ is given by $\mathit{SINR}_{ij}(t)=x_{ij}(t)p_{ij}(t)g_{ij}(t)/(1+I_{ij}(t))$, where $g_{ij}(t)=|h_{ij}(t)|^2$ is the channel power gain, which is normalized to get a noise power of $1$. The term $I_{ij}(t)$ denote the power of the interference coming from other devices and transmitting using RB $(j,t)$. 

To accommodate a large number of IoT devices, power-domain NOMA is used in this paper, where a group of IoT devices transmit to the gNB over the same RB. For decoding, successive interference cancellation (SIC) is used at the gNB. Let $\mathbb{A}_{j}(t)$ denote the set of IoT devices transmitting using RB $(j,t)$. It is well-know that the highest channel decoding order is used in uplink NOMA~\cite{8632657}. That is, the interference received at the gNB, which is generated by the transmission of device $i$'s packet, comes from all devices that have lower channel gains. Let $\mathbb{B}_{ij}(t)\coloneq\{i'\in\mathbb{A}_{j}(t):g_{i'j}(t)<g_{ij}(t)\}$ be the ordered set of devices with respect to uplink NOMA. With that said, the interference received by the gNB can be calculated as $I_{ij}(t)=\sum_{i'\in\mathbb{B}_{ij}(t)}x_{i'j}(t)p_{i'j}(t)g_{i'j}(t)$. The achievable rate (in bits/s) between device $i$ and the gNB using RB $(j,t)$ is given by $R_{ij}(t)=W\lg(1+\text{SINR}_{ij}(t))$.

The considered problem is called NOMA grouping and power allocation (NG-PA). The objective function of NG-PA is to maximize the number of successfully delivered packets by each IoT device during the time horizon of $T$ frames. This has to be done subject to NOMA grouping and power allocation constraints. Solving NG-PA is done in an online fashion in which each IoT device only knows the current and previous information and devices may communicate with each other through the gNB using feedback and uplink channels. NG-PA can be written as the following integer linear program.
\begin{problem}\label{pb:1}
  \begin{alignat}{2}
  & \maximize &\quad &\sum_{i=1}^M\sum_{j=1}^N\sum_{t=1}^Tx_{ij}(t)\label{obj:1}\\
  & \subjto
  & & x_{ij}(t)\in\{0,1\},p_{ij}(t)\in\mathbb{P}\label{cns1:1}\\
  & & & \sum_{j=1}^NR_{ij}(t)\geqslant\sum_{j=1}^NL_{i}(t)x_{ij}(t)\label{cns2:1}\\
  & & & p_{ij}(t)\leqslant p_{i}^{\textsf{max}}x_{ij}(t)\label{cns3:1}\\
  & & & \sum_{j=1}^N\sum_{t=1}^T p_{ij}(t)\leqslant p_{i}^{\textsf{max}}\label{cns4:1}\\
  & & & x_{ij}(t) = 0, \forall j\notin\{a_{i}(t)..d_{i}(t)-1\}\label{cns5:1}\\
  & & & \sum_{i=1}^Mx_{ij}(t)\leqslant G, \text{(P1g)},\sum_{j=1}^Nx_{ij}(t)\leqslant 1\tag{P1h}
  \end{alignat}
\end{problem}
The objective function in~\eqref{obj:1} maximizes the number of successfully delivered packets  during $T$ frames. Constraints~\eqref{cns1:1} list the optimization variables where $p_{ij}(t)$ belongs to the discrete set $\mathbb{P}$. Constraints~\eqref{cns2:1} guarantee the minimum requirements of device $i$. Constraints~\eqref{cns3:1} and~\eqref{cns4:1} guarantee the limited operating energy of device $i$. Constraints~\eqref{cns5:1} respect the arrival and deadline of device $i$. Constraints~(P1g) and~(P1h) guarantee at most $G\leqslant M$ devices per RB and at most a RB per device, respectively. The one RB per device is realistic in massive access problem since devices have short packets~\cite{7165674}. 

Since each device will be allocated some amount of transmission power in each frame, thus, if such amount could be known, one could reduce the problem to $T$ single-frame subproblems and solve each one separately. For this purpose, we start by analyzing NG-PA in the case of one frame (the superscript of $t$ is dropped when not needed).

\section{The Competitive Reinforcement Learning Framework}\label{sec:learn}
\subsection{A Competitive Algorithm}
Here, we focus on frame $t$ and we assume that the transmission power of device $i$ at $t$ is known and we solve sub-optimally the NOMA grouping (NG) subproblem. Device $i$ can choose its transmission power from $\{0,p_i\}$. 

To solve NG, we transform it into a many-to-one matching problem and we adopt a greedy approach. We create the following bipartite graph. The devices represent the right vertexes and the slots represent the left vertexes that appear in an online fashion. An edge exists between slot $j$ and device $i$ if and only if $p_ig_{ij}\geqslant(2^{L_i/W}-1)$ and $j\in\{a_i..d_i-1\}$. When slot $j$ appears, the channel gain $g_{ij}$ is revealed for all devices $i$ and thus the edges incident to it are also revealed. Once revealed, an online algorithm must make an irrevocable decision of which device to serve at slot $j$ (i.e., match the corresponding edge). Now, each slot can be matched to at most $G$ devices from those connected to it by an edge. For each slot $j$, let $\mathbb{N}_j$ denotes the set of neighbors of $j$ (i.e., $\mathbb{N}_j\coloneq\{i:\{i,j\} \text{ is an edge}\}$). Once slot $j$ is revealed, the problem is reduced to  finding a set of (at most $G$) devices $\mathbb{D}_j\subseteq\mathbb{N}_j$ of maximum cardinality such that:
\begin{align}\label{binslot}
p_ig_{ij}\geqslant(2^{L_i/W}-1)(1+\sum_{i'\in\mathbb{D}_j'}p_{i'}g_{i'j}),\forall i\in\mathbb{D}_j,
\end{align}
where $\mathbb{D}_j'\coloneq\{i'\in\mathbb{D}_j:g_{ij}>g_{i'j}\}$.


The proposed algorithm to solve NG is called frame-matching (\textsc{fm}) and its pseudo-code is given in Algorithm~\ref{alg:onmatchingM}. For each arriving slot, \textsc{fm} applies a greedy approach to match each device to each new arriving slot---starting with the device with the lowest channel gain. The greedy approach gives a maximum cardinality set that satisfies~\eqref{binslot} in $\mathscr{O}(M\lg M)$ time in the worst-case. (See~\cite{mlika2020massive} for a complete and detailed proof.) The worst-case time complexity of \textsc{fm} is clearly $\mathscr{O}(NM\lg M)$. \textsc{fm} serves the maximum possible number of devices in each slot. Then, it updates the set of not-yet-served devices and continues in this way for the next slot. Line 7 of the \textsc{fm} algorithm just makes sure that the size of $\mathbb{D}_j$ does not exceed the NOMA size $G$, i.e., we select any subset of $\mathbb{Y}$ with size $G$.\textsc{fm} returns the list of grouped devices in each slot $\{\mathbb{D}_1,\mathbb{D}_2,\ldots,\mathbb{D}_N\}$ as well as the total number of served devices $\sum_{j=1}^N|\mathbb{D}_j|$.
\begin{algorithm}[ht!]
  \caption{The \textsc{fm} algorithm}
  \label{alg:onmatchingM}
  \begin{algorithmic}[1]
    \Require{Bipartite graph, $G,M,N,[g_{ij}],[L_i],[p_i]$}
    \Ensure{$\{\mathbb{D}_1,\mathbb{D}_2,\ldots,\mathbb{D}_N\}$}
    \State $\mathbb{D}\gets\{1,2,\ldots,M\}$
    \For{$j\in\{1,2,\ldots,N\}$}
        \State $\mathbb{Y}\gets\emptyset$; $Y\gets0$
        \For{$i\in\mathbb{N}_j$}
            \If{$p_ig_{ij}\geqslant (2^{L_i/W}-1)(1+Y)$}
                    \State $\mathbb{Y}\gets\mathbb{Y}\cup\{i\}$; $Y\gets Y+p_ig_{ij}$
              \EndIf  
        \EndFor
        \State Let $\mathbb{D}_j\subseteq\mathbb{Y}$ with $|\mathbb{D}_j|\leqslant G$
        \State $\mathbb{D}\gets\mathbb{D}\backslash\mathbb{D}_j$ and find $\mathbb{N}_j$
    \EndFor
    \State\Return $\{\mathbb{D}_1,\mathbb{D}_2,\ldots,\mathbb{D}_N\}, \sum_{j=1}^N|\mathbb{D}_j|$
  \end{algorithmic}
\end{algorithm}


\subsection{A Competitive Reinforcement Learning Algorithm}

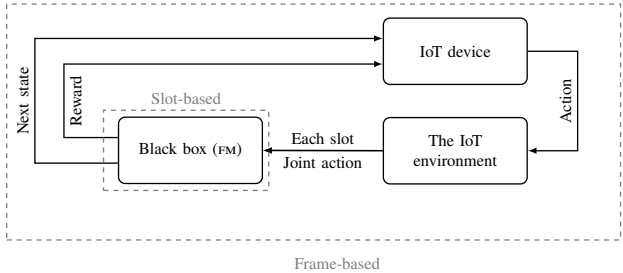
\begin{figure}[htpb!]
  \centering
  \resizebox{.5\textwidth}{!}{%
\begin{tikzpicture}[node distance = 6em, auto, thick]
    \node [rectangle,draw,text width=8em,text centered,rounded corners,minimum height=4em] (Agent) {IoT device};
    \node [rectangle,draw,text width=8em,text centered,rounded corners,minimum height=4em, below of=Agent] (Environment) {The IoT environment};
    \node [rectangle,draw,text width=8em,text centered,rounded corners,minimum height=4em, left of=Environment] at (-3.5, -2.1) (BlackBox) {Black box (\textsc{fm})};
    
     \path [draw, -latex] (Agent.0) --++ (3em, 0em) |- node [above,pos=0.25,rotate=90]{Action} (Environment.0);
     \path [draw, -latex] (Environment) --++ (BlackBox) node [midway, above]{Each slot};
     \path [draw, -latex] (Environment) --++ (BlackBox) node [midway, below]{Joint action};
     \path [draw, -latex] (BlackBox.190) --++ (-5em,0em) |- node [above,pos=0.25,rotate=90] {Next state} (Agent.170);
     \path [draw, -latex] (BlackBox.170) --++ (-3.25em,0em) |- node [below,pos=0.25,rotate=90] {Reward} (Agent.190);
     \node (rect) at (-5.7,-2.1) [draw,dashed,color=gray,thick,minimum width=3.5cm,minimum height=1.7cm] {};
     \node[gray,thick] at (-5.7,-1) {Slot-based};
     \node (rect) at (-3,-1.5) [draw,dashed,color=gray,thick,minimum width=13cm,minimum height=5cm] {};
     \node[gray,thick] at (-2.5,-4.5) {Frame-based};
\end{tikzpicture}
  }
  \caption{The system block of the CRL framework.}
  \label{fig:block}
\end{figure}

To solve solve NG-PA efficiently, we propose a CRL framework as illustrated in Fig.~\ref{fig:block}. Each agent (or, interchangeably, device), interacts independently with the IoT environment and takes actions accordingly. The learning is a frame-based process. In each frame $t$, each device $i$ observes the IoT environment and chooses a transmission power $p_i(t)$ from its available set of actions $\mathbb{P}$. Once all devices choose their transmission powers, a joint action is formed and a slot-based process is invoked---the \textsc{fm} algorithm---as a black box. Just before the beginning of the next frame $t+1$, the number of successfully received packets in frame $t$ is calculated by the gNB using \textsc{fm} and a reward signal is obtained. The gNB broadcasts this reward signal to each agent and the IoT environment moves to the next state. The reward given by \textsc{fm} is common to all agents to incite a cooperative behavior among devices. Thanks to the simplicity of \textsc{fm}, our approach solves perfectly the curse of dimensionality issue in RL. The details of this learning framework process is given in the sequel.

We model NG-PA as an online deterministic multi-agent Markov decision process (MDP). This modeling helped us to transform NG-PA to an online stochastic shortest path problem. The MDP is deterministic because the transition probabilities are known. The multi-agent MDP can be seen as multiple MDPs---one for each agent. The corresponding transition graph (TG) of each MDP is constructed as follows. A state in each TG $i$ (corresponding to agent $i$) is given by $(e_{i}(t),t)$ where $e_i(t)$ is the remaining energy level at frame $t$ in device $i$'s battery. For any state $(e_{i}(t),t)$ of TG $i$, an action is given by the transmission power $p_{i}(t)\in\mathbb{P}$. When $t=1$, the node $\mathbf{s}_i\coloneq(e_i(1),1)$ is called the starting node with $e_{i}(1)=p_i^{\textsf{max}}$ for all $i$. There is a terminal node denoted by $\mathbf{t}_i\coloneq(e_i(T+2),T+2)$ with $e_{i}(T+2)=0$ for all $i$. For $t=1,2,\ldots,T+1$, a transition from $(e'(t),t)$ to $(e''(t+1),t+1)$ happens with probability one if and only if $e'(t)-e''(t+1)\geqslant0$. No other transition is allowed. The possible actions in state $(e(t),t)$ are given by the outgoing edges of node $(e(t),t)$. 

Normally, when device $i$, in state $(e_{i}(t),t)$, chooses action $p_{i}(t)$, its reward is a binary number that represents whether or not it is served. Designing the rewards in this way teaches the devices to act selfishly and thus does not necessarily give good outcome, i.e., the objective function could be very low because each device will learn to use its transmission power to get served regardless of others. It is thus necessary to redesign the rewards to improve the learning outcome. The idea of our CRL framework comes from this important remark. Thus, instead of the binary rewards, each device receives its reward from the black box---the \textsc{fm} algorithm---in each frame. This can be acquired by information feedback between the devices and the gNB. The proposed \textsc{crl} algorithm works as follows.

Each device learns its own $\mathbf{s}$-$\mathbf{t}$ path by applying a modified version of \textsc{exp3}~\cite{10.1137/S0097539701398375}---a popular RL algorithm for the adversarial multi-armed bandit problem based on exponential-weighting for exploration and exploitation. \textsc{crl} operates in rounds, where in each round, it is applied at device $i$ that chooses an $\mathbf{s}_i$-$\mathbf{t}_i$ path according to some probability (proportional to the path weight). This probability is chosen to follow a distribution over the set of all $\mathbf{s}_i$-$\mathbf{t}_i$ paths in order to get a mixture between exponential weighting of biased estimates of the rewards and uniform distribution to ensure sufficiently large exploration of each edge of any $\mathbf{s}_i$-$\mathbf{t}_i$ path. After choosing an $\mathbf{s}_i$-$\mathbf{t}_i$ path, device $i$ gets to know the rewards on each edge of that path, i.e, it gets to know the number of successfully received packets in the corresponding frame. Then, \textsc{crl} updates the probability distribution (by updating the paths weights) and continues similarly. Every TG $i$ has $2+TP$ nodes with $P\coloneq|\mathbb{P}|$ and $P(P(T-1)+T+3)/2$ directed edges. Every path in TG $i$ has length $T+1$. Let $\mathbb{P}_i$ be the set of all $\mathbf{s}_i$-$\mathbf{t}_i$ paths in TG $i$ and let $\sigma_i\coloneq|\mathbb{P}_i|$. We can prove that $\sigma_i=\binom{T+P-1}{T}$, which is exponentially large and thus choosing the paths in this way according to their weights is not efficient. However, a simple modification can improve the algorithm enormously. First, instead of assigning weights to paths, they are assigned to edges. Second, we construct a set of edge-covering $\mathbf{s}_i$-$\mathbf{t}_i$ paths $\mathbb{C}_i$, which is defined as the set of paths in TG $i$ such that for any edge $e$ in TG $i$, there is a path $\mathbf{p}_i$ in $\mathbb{C}_i$ such that $e\in\mathbf{p}_i$. Such an edge-covering paths $\mathbb{C}_i$ can be obtained in $\mathscr{O}(TP^2+TP\lg(TP))$ time using Dijkstra's algorithm where $|\mathbb{C}_i|=\mathscr{O}(TP^2)$. Now, instead of each path, each edge $e$ of TG $i$ is assigned a weight $w(e)$ (initialized to one for each edge at the beginning of the round) and the weight of an $\mathbf{s}_i$-$\mathbf{t}_i$ path is given by the product of the weights of its edges. For each round, \textsc{crl}, applied at device $i$, chooses an $\mathbf{s}_i$-$\mathbf{t}_i$ path (1) uniformly from $\mathbb{C}_i$ with probability $\gamma$ or (2) according to the paths weights with probability $1-\gamma$. If the latter is to be done, then the $\mathbf{s}_i$-$\mathbf{t}_i$ path can be chosen by adding its vertexes one-by-one according to edges' weights (and not to paths' weights). Next, \textsc{crl} finds the probability of choosing each edge in the TG $i$, which can also be done using edges' weights only based on paths kernels and dynamic programming. Then, for each frame (or edge), the rewards are obtained using \textsc{fm}, where the reward $r$ at any edge is normalized by the probability of that edge $q(e)$, i.e., the normalized reward is $(\beta+r\mathds{1}_{\{e\in\mathbf{p}_i\}})/q(e)$, with $\mathds{1}_{\mathbb{A}}$ denotes the indicator function and $\beta\in(0,1]$. Finally, the edges' weights are updated as $w(e)\gets w(e)e^{\eta r}$ where $\eta>0$.

The per-round complexity of \textsc{crl} is given by $\mathscr{O}(MTP^2+TNM\lg M)$, where $\mathscr{O}(TNM\lg M)$ is the complexity of applying \textsc{fm} in all frames and $\mathscr{O}(MTP^2)$ is the complexity of choosing the paths according to the edges' weights and updating the probability of each edge.

\begin{figure*}[!b]
	\centering
	\begin{minipage}[b]{.4\textwidth}
		\includegraphics[scale=0.43]{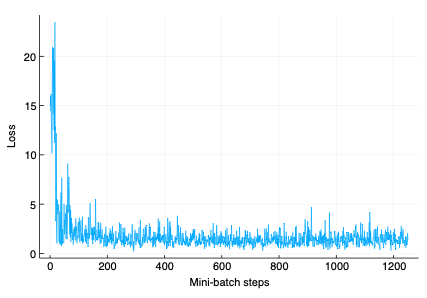}
		\caption{The training loss.} 
		\label{dql:1}
	\end{minipage}\qquad
	\begin{minipage}[b]{.4\textwidth}
		\includegraphics[scale=0.43]{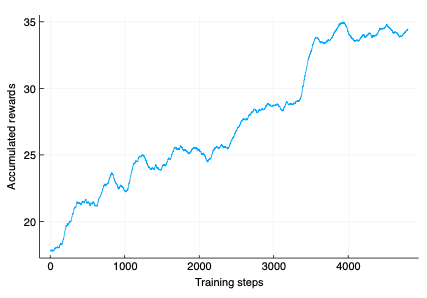}
		\caption{The cumulative reward.} 
		\label{dql:2}
	\end{minipage}
\end{figure*}

\section{Benchmark DRL Algorithm}\label{SubSec1}
We implemented the off-the-shelf method of DQL (\textsc{dql}) as a benchmark solution to compare our proposed \textsc{crl}. \textsc{dql} is proposed in~\cite{9318243} to solve the resource allocation problem in multiaccess edge computing-based Internet of vehicles network using network slicing. It uses deep neural networks to approximate the Q function. We use prioritized replay memory, double Q and dueling architecture to help agents remember and use past experiences. \textsc{dql} is implemented in the multi-agent scenario in which each agent (IoT device) trains its own deep-Q-network (DQN). Each DQN is associated some weight vector to represent the Q function. The input to each DQN is given by the observation of the current state. The output is the value of the Q function which is given by the appropriate action taken by each agent.

The training goes in episodes where each episode lasts $T$ frames, that is the training is a frame-based process. For any episode, the training starts at frame $1$ (the initial state) and finish at frame $T$ (the terminal state). For an agent, a state is given by (1) the channel gains between the agent and the gNB across all slots, (2) an indicator vector of length $M$ to indicate which agent is served in the current frame, (3) the remaining energy level of the agent, (4) the arrival time and the deadline of the agent, (5) the frame and the episode index, and (6) the current exploration rate. At any episode, the possible actions of an agent in some state are given by all pairs of slots and power levels. That is, in frame $t$, if the remaining energy level of device $i$ is $e_i(t)$, then the possible action is $(j,p_i)\in\{1,2,\ldots,N\}\times\mathbb{P}$ with $p_i\leqslant e_i(t)$. The agent uses the $\epsilon$-greedy policy to explore the action space. Once all agents choose their actions, the gNB calculates the number of successfully delivered packets and each agent moves to the next state (frame $t+1$). The reward of each agent is incremented in each frame until frame $T$ to count the overall number of successfully delivered packets. The tuple of (state, action, next state, reward) is stored in the prioritized replay memory with some associated priority. After some episodes, a mini-batch of stored experiences is sampled from the prioritized replay memory according to the associated priorities. This mini-batch is used to update the weight parameter of each DQN using a variant of the stochastic gradient descent algorithm in order to minimize the loss function. The loss function is given by the mean-squared error of the discounted reward and the value of the Q function so far. To calculate the loss function, a duplicate copy of the original DQN (the target DQN) is created in order to update the original DQN once in a while. The exploration rate $\epsilon$ is annealed based on the episode index. Annealing the exploration rate over time is a technique used in RL to solve the dilemma between exploration and exploitation, i.e., as the time goes by, we decrease $\epsilon$ to increase the exploitation probability as the agent starts to learn something useful.

We note that this \textsc{dql} algorithm has more knowledge than the proposed \textsc{crl} as the former knows all information in the current frame but the latter knows only the information in the current slot. Besides, \textsc{dql} requires training-intensive tasks in which the DQNs should be always trained and the hyperparameters should be tunned carefully to achieve the best performance.



\begin{figure*}
  \centering
  \begin{minipage}[b]{.4\textwidth}
  \captionsetup{justification=centering,margin=2cm}
  \resizebox{1\textwidth}{!}{%
    \begin{tikzpicture}
    \tikzset{every pin/.style={draw=black,fill=yellow!20,rectangle,rounded corners=3pt,font=\scriptsize},}
    \begin{axis}[
    xlabel={Maximum packet length $L_{\textsf{max}}$ (in bits)},
    ylabel={Avg. number of successfully delivered packets},
    set layers,
    grid=both,
    legend cell align=left,
    xmin=200,
    xmax=500,
    ymin=15,
    ymax=35,
    xtick={200,300,400,500},
    x label style={font=\footnotesize},
    y label style={font=\footnotesize}, 
    ticklabel style={font=\footnotesize},
    legend style={at={(.97,.86)},anchor=east,font=\scriptsize},
    ]
    \addplot[dashed,color=black,mark=asterisk,mark options={scale=1,solid}] coordinates {
      (200, 32.7) (300, 29.83) (400, 27.81) (500, 26.54)
    };\addlegendentry{\textsc{opt}}
    \addplot[dashed,color=blue,mark=o,mark options={scale=1,solid}] coordinates {
      (200, 24.32) (300, 23.53) (400, 22.6) (500, 21.92)
    };\addlegendentry{\textsc{crl}}
    \addplot[dashed,color=red,mark=triangle,mark options={scale=1,solid}] coordinates {
      (200, 21.15) (300, 20.24) (400, 19.51) (500, 18.5)
    };\addlegendentry{\textsc{dql}}        
    \addplot[dashed,color=violet,mark=diamond,mark options={scale=1,solid}] coordinates {
      (200, 18.53) (300, 17.96) (400, 17.44) (500, 17.078)
    };\addlegendentry{\textsc{tql}}
    \end{axis}
    \end{tikzpicture}
  }
  \caption{Impact of packet sizes on \textsc{crl}.}
  \label{crl:1}
\end{minipage}\qquad
\begin{minipage}[b]{.4\textwidth}
  \captionsetup{justification=centering,margin=2cm}
  \resizebox{1\textwidth}{!}{%
    \begin{tikzpicture}
    \tikzset{every pin/.style={draw=black,fill=yellow!20,rectangle,rounded corners=3pt,font=\scriptsize},}
    \begin{axis}[
    xlabel={Group size ($G$)},
    ylabel={Avg. number of successfully delivered packets},
    set layers,
    grid=both,
    legend cell align=left,
    xmin=2,
    xmax=20,
    ymin=15,
    ymax=40,
    xtick={2,8,14,20},
    x label style={font=\footnotesize},
    y label style={font=\footnotesize}, 
    ticklabel style={font=\footnotesize},
    legend style={at={(.23,.86)},anchor=east,font=\scriptsize},
    ]
    \addplot[dashed,color=black,mark=asterisk,mark options={scale=1,solid}] coordinates {
      (2, 32.7) (8, 33.57) (14, 34.21) (20, 35.15)
    };\addlegendentry{\textsc{opt}}
    \addplot[dashed,color=blue,mark=o,mark options={scale=1,solid}] coordinates {
      (2, 24.32) (8, 24.47) (14, 24.94) (20, 25.24)
    };\addlegendentry{\textsc{crl}}
    \addplot[dashed,color=red,mark=triangle,mark options={scale=1,solid}] coordinates {
      (2, 21.15) (8, 21.6) (14, 21.77) (20, 21.92)
    };\addlegendentry{\textsc{dql}}        
    \addplot[dashed,color=violet,mark=diamond,mark options={scale=1,solid}] coordinates {
      (2, 18.53) (8, 19.07) (14, 19.7) (20, 19.77)
    };\addlegendentry{\textsc{tql}}
    \end{axis}
    \end{tikzpicture}
  }
  \caption{Impact of group sizes on \textsc{crl}.}
  \label{crl:2}
\end{minipage}
\end{figure*}

\section{Simulation Results}\label{sec:simulations}
This section illustrates the performance of the proposed \textsc{crl} framework by comparing it to \textsc{dql} and to an optimal offline solution \textsc{opt} implemented based on~\eqref{pb:1} using off-the-shelf solvers. We consider a dense geographical zone modeled by a square of side $20$ meters in which the gNB is located at the center and $M=20$ IoT devices are randomly and uniformly distributed inside the square. The simulations parameters are based on 3GPP specifications~\cite[p.~481]{3gpp.45.820} and are given as follows. The carrier frequency is $900$ MHz, $W=40$ kHz, path-loss follows $120.9+37.6\log(\text{dist}_{i}(t))+\alpha_{\text{G}}+\alpha_{\text{L}}$, where $\text{dist}_{i}(t)\in[0,0.02]$ is the distance in km, the antenna gain  $\alpha_{\text{G}}=-4$ dB, the penetration loss $\alpha_{\text{L}}=10$ dB, Rayleigh fading is considered, the noise figure is $5$ dB, $\mathbb{P}=\{-100,17,21,23\}$, $G=2$, $L_i(t)\sim\mathrm{unif}\{100, 500\}$ kbits, $a_i(t)\sim\mathrm{unif}\{1,N\}$, $d_i(t)\sim\mathrm{unif}\{a_i(t)+1,N+1\}$, and the noise power is $-174$ dBm/Hz

For comparison purposes, we also implemented the tabular Q learning (\textsc{tql}) algorithm with learning rate of $\alpha=0.5$. We train $M=20$ DQNs with $N=5$ slots and $T=5$ frames. The DQNs are created and trained in the Julia programming language using the machine learning library Flux.jl~\cite{innes:2018}. Each DQN consists of an input and an output layers and of three fully connected hidden layers containing respectively $50$, $35$, and $20$ neurones. The activation function rectified linear unit (ReLU) is used in each layer. Each DQN is trained with the RMSProp optimizer with a learning rate of $5*10^{-3}$. The training lasts $5000$ episodes with an exploration rate starting from $0.2$ and annealed to reach $0.01$. The target update frequency is $10$ episodes and the mini-batch training frequency is each episode. The mini-batch size is chosen equal to $300$.
The \textsc{crl}'s parameters are $\gamma=0.5$, $\beta=0.01$, and $\eta = 0.00075$ and the number of rounds is $50$. 

Figs.~\ref{dql:1} and~\ref{dql:2} present the cumulative reward (averaged over the last $200$ episodes) of all agents versus the episodes as well as the loss achieved by a single (randomly chosen) agent versus the mini-batch steps (a mini-batch training is done every $4$ episodes). The cumulative reward improves as the training episodes increase and the loss decreases to reach a value close to zero.

Figs.~\ref{crl:1} and~\ref{crl:2} present the performance of \textsc{crl} for different values of packets sizes and NOMA group sizes and they compare it to \textsc{dql}, \textsc{tql} and \textsc{opt}, where the latter is obtained through solving~\eqref{pb:1} using off-the-shelf solvers. We can see that our proposed CRL framework gives superior results despite being online and executed with only few rounds ($50$ rounds). However, the training-intensive \textsc{dql} algorithm is trained for huge number of episodes and gives inferior results. It might be possible to improve the results of \textsc{dql} by tunning further the hyperparameters but doing so will only increase the complexity and the overhead. Since the optimal algorithm is omniscient it has the highest performance, e.g., in Fig.~\ref{crl:2}, \textsc{crl} is $27\%$ less than \textsc{opt}.

\section{Conclusion}\label{sec:conclusions}
In this paper we studied NOMA grouping and power allocation in IoT networks. To solve the problem in a practical way, we divided it into NOMA grouping subproblems. Then, we proposed online competitive algorithms (CAs) to solve the subproblems. To obtain the transmission power allocation solution, we proposed a competitive-assisted reinforcement learning (CRL) framework that uses the CAs. We showed that the proposed CRL framework, without requiring training-intensive tasks, achieves superior performance and beats off-the-shelf DRL methods such as DQL.
  

\bibliographystyle{IEEEtran}
\bibliography{IEEEabrv,icc}

\end{document}